\documentclass[twocolumn, switch]{article} 
\usepackage{preprint}

\usepackage{amsmath, amsthm, amssymb, amsfonts}

\usepackage[numbers,square]{natbib}
\usepackage[utf8]{inputenc}	
\usepackage[T1]{fontenc}	
\usepackage{xcolor}		
\usepackage[colorlinks = true,
            linkcolor = purple,
            urlcolor  = blue,
            citecolor = cyan,
            anchorcolor = black]{hyperref}	
\usepackage{booktabs} 		
\usepackage{nicefrac}		
\usepackage{microtype}		
\usepackage{lineno}		
\usepackage{float}			

\usepackage{lipsum}		

\usepackage{newfloat}
\DeclareFloatingEnvironment[name={Supplementary Figure}]{suppfigure}
\usepackage{sidecap}
\sidecaptionvpos{figure}{c}

\usepackage{titlesec}
\titlespacing\section{0pt}{12pt plus 3pt minus 3pt}{1pt plus 1pt minus 1pt}
\titlespacing\subsection{0pt}{10pt plus 3pt minus 3pt}{1pt plus 1pt minus 1pt}
\titlespacing\subsubsection{0pt}{8pt plus 3pt minus 3pt}{1pt plus 1pt minus 1pt}

\title{Orbital Graph Convolutional Neural Network for Material Property Prediction}

\usepackage{xwatermark}
\newwatermark[firstpage,color=gray!60,angle=90,scale=0.32, xpos=-4.05in,ypos=0]{\href{https://doi.org/}{\color{gray}{Publication doi}}}
\newwatermark[firstpage,color=gray!60,angle=90,scale=0.32, xpos=3.9in,ypos=0]{\href{https://doi.org/}{\color{gray}{Preprint doi}}}
\newwatermark[firstpage,color=gray!90,angle=0,scale=0.28, xpos=0in,ypos=-5in]{†correspondence: \texttt{barati@cmu.edu} \\{* Equal Contribution}}

\usepackage{authblk}

\author[1,2,*]{Mohammadreza Karamad}
\author[1,*]{Rishikesh Magar}
\author[1]{Yuting Shi}
\author[3]{Samira Siahrostami}
\author[2]{Ian D. Gates}
\author[1,†]{Amir Barati Farimani}

\affil[1]{Department of Mechanical Engineering, Carnegie Mellon University}
\affil[2]{Department of Chemical and Petroleum Engineering, University of Calgary}
\affil[3]{Department of Chemistry,University of Calgary}
\begin{document}

\twocolumn[ 
  \begin{@twocolumnfalse} 
  
\maketitle

\begin{abstract}
Material representations that are compatible with machine learning models play a key role in developing
models that exhibit high accuracy for property prediction. Atomic orbital interactions are one of
the important factors that govern the properties of crystalline materials, from which the local chemical
environments of atoms is inferred. Therefore, to develop robust machine learningmodels for material
properties prediction, it is imperative to include features representing such chemical attributes. Here,
we propose the Orbital Graph Convolutional Neural Network (OGCNN), a crystal graph convolutional
neural network framework that includes atomic orbital interaction features that learns material
properties in a robust way. In addition, we embedded an encoder-decoder network into the OGCNN
enabling it to learn important features among basic atomic (elemental features), orbital-orbital interactions,
and topological features. We examined the performance of this model on a broad range
of crystalline material data to predict different properties. We benchmarked the performance of the
OGCNN model with that of: 1) the crystal graph convolutional neural network (CGCNN), 2) other
state-of-the-art descriptors for material representations including Many-body Tensor Representation
(MBTR) and the Smooth Overlap of Atomic Positions (SOAP), and 3) other conventional regression
machine learning algorithms where different crystal featurization methods have been used. We find
that OGCNN significantly outperforms them. The OGCNN model with high predictive accuracy can
be used to discover new materials among the immense phase and compound spaces of materials.
\end{abstract}
\vspace{0.35cm}

  \end{@twocolumnfalse} 
] 



\section{Introduction}
Owing to the methodological improvements in ab initio calculations, such as density functional theory (DFT) as well as increasing computing power, it has now become possible to perform high-throughput computational calculations to search for new materials with specific properties of interest \cite{Franceschetti1999TheIB,Ceder1998,PhysRevLett.88.255506,C2EE22341D,PhysRev.136.B864}. However, ab initio high-throughput computational methods are hampered by expensive calculations necessitating development of alternative methods to predict material properties. Machine Learning (ML) techniques, on the other hand, have proven to provide a fast and accurate way to predict desired properties enabling facile discovery of new materials at a fraction of the computational cost and in a shorter time scale. ML  algorithms  build a functional map between the input data representing the material and the output data being the properties of interest. These models have been used to predict a wide range of properties for different classes of material \cite{PhysRevLett.114.105503,PhysRevLett.115.205901, PhysRevB.93.085142, PhysRevB.89.094104, DEY2014185, Xue2016, Isayev2017, ZhouE6411, doi:10.1021/acs.jpclett.5b01660,doi:10.1021/acs.chemmater.7b00156,LIU2017159}.
One of the important challenges to develop a ML-based approach for predicting material properties is material representation, i.e. encoding material information, including features (also often called the descriptors), geometrical and topological information \cite{PhysRevB.87.184115}.The features need to be unique in representing material, they should be computed at low computational cost or preferably be readily accessible from available databases. Most importantly, they should reflect the chemical information related to the targeted properties. This requires encoding the information about electronic structure, chemistry as well as the topology of the material. In addition, feature vector representation needs to be compatible with the ML model. To this end, developing features that possess the aforementioned properties has proven to be challenging \cite{PhysRevLett.114.105503}. \\
Roughly speaking, for a given crystalline material, the information related to its physical and chemical properties arise from the position of charges and nuclei, the topology of the crystal, basic properties of its constituent elements, and interatomic interactions. Furthermore, key information about the local chemical environments of atoms forms the fundamental basis to determine the properties of the crystals. Therefore, the accuracy of a ML model to predict material properties is mostly controlled by the ability of its descriptors to accurately encode the local chemical environments of atoms \cite{PhysRevB.87.184115}. Different methods to represent the local chemical environments of atoms have been developed. Examples include using atom-distribution-based symmetry functions \cite{PhysRevLett.98.146401,doi:10.1063/1.3553717}, Smooth Overlap of Atomic Positions\cite{PhysRevB.87.184115}, Many-body Tensor Representation \cite{huo2017unified}, band structures and density of states descriptors \cite{doi:10.1021/cm503507h}, and Coulomb matrix representation \cite{PhysRevLett.108.058301}. In many of these methods either structural or elemental information or both have been used as features for representation. The elemental features are either intrinsic quantities such as the atomic number and ionization energy or heuristic quantities such as the electronegativity and ionic radius \cite{Ward2016, doi:10.1021/cm503507h, Ye2018}. Structural representations, on the other hand, encode local chemical environments by capturing the geometry and interaction between atoms \cite{PhysRevB.89.205118, PhysRevLett.108.058301}. Another interesting attempt for crystal representation uses electronic structure attributes \cite{doi:10.1021/cm503507h, PhysRevB.89.094104}. This is important because the electronic structure is one of the key parameters in defining its properties. The electron configurations and orbital-orbital interactions are important electronic structure attributes which should be included as representations for ML predictions. Including such information, however, often requires performing DFT calculations to generate descriptors which consequently increases computational costs \cite{doi:10.1021/cm503507h}.
Different efforts have been made to represent material by considering electronic structure attributes without performing DFT calculations \cite{Ward2016, doi:10.1080/14686996.2017.1378060}.
Ward et al. used an extensive set of features, including basic atomic and electronic structure features to develop a ML model for predicting different crystalline properties  \cite{Ward2016, PhysRevB.89.094104}. They used the average fraction of electrons from the s, p, d and f valence shells of all present elements to quantitatively represent atomic electronic states as electronic structure attributes. Clearly, such representations do not explicitly include orbital interactions among constituent elements of the crystals. In another work, Pham et al. developed a novel two-dimensional descriptor called the orbital field matrix (OFM) that encodes orbital interactions according to electron configurations of central atom and neighbor atoms surrounding the central atom \cite{doi:10.1080/14686996.2017.1378060}.
In OFM, a simple description of the interaction of valence electrons of a central atom with its neighbor atoms represents orbital-orbital interactions. The OFM model showed promise to predict different material properties including formation energy and atomization energy by using conventional ML algorithms such as kernel ridge regression, decision tree regression, and random forest regression. Despite its promise, the OFM model does not include any elemental atomic features, kernelized distance features, or graph representation of the crystals. In addition, OFM model did not use the state of the art deep learning techniques.\\
Deep learning has been widely used in materials science research and molecular property prediction \cite{gomes2017atomic, Back2019,doi:10.1021/acs.jcim.9b00550,C7SC02664A,LI2017232}. In particular, Convolutional Neural Networks (CNNs) have been used for material properties prediction because of their special ability to extract features from the data \cite{10.5555/2969442.2969488}. In a recent study by Cao et al., Magpie and OFM descriptors were used in conjunction with CNN to predict material properties \cite{Cao2019}. The reported prediction accuracy for training formation energies of an alloy dataset is significantly higher than using either of descriptors. However, Cao et al. did not use graph representation for crystalline systems, feature representation, and dimensionality reduction. Recently, Xie et al., developed a crystal graph convolutional neural network (CGCNN) framework to represent periodic crystalline systems for predicting material properties \cite{PhysRevLett.120.145301}.
In CGCNN, graph representation was used to describe the structure of the crystals. In addition, the crystal information was encoded using basic atomic features such as electron affinity and group number. To encode the neighboring atoms geometrical effects, the interatomic interactions using their atomic distances  were considered. The CGCNN model has many powerful characteristics such as inclusion of kernelized distance features, features encoding via dimensionality reduction, and convolution of the atom features with its neighbors. Although the CGCNN model demonstrates ability to predict a variety of properties with high accuracy, it does not consider the attributes that contain orbital-orbital interaction features. In this contribution, we develop a graph convolutional neural network that incorporates the atomic orbital interactions. This new model is referred to as the Orbital Graph Convolutional Neural Network (OGCNN). The inclusion of orbital-orbital interactions to encode the local chemical environments of atoms, along with embedding of an encoder-decoder network enabled OGCNN to achieve higher accuracy compare to CGCNN. To show the robustness of the OGCNN model, we benchmarked it against CGCNN, other state-of-the-art descriptors for material representations including Many-body Tensor Representation (MBTR) and the Smooth Overlap of Atomic Positions (SOAP), and a variety of conventional ML models with different crystalline representations.

\section{Model Architecture}
To include atomic orbital interactions, we employed the representation of crystal systems named OFM where the atomic orbital interactions are counted based on the distribution of valence shell electrons \cite{doi:10.1080/14686996.2017.1378060,doi:10.1063/1.5021089}. In the OFM model, the electron configuration of each atom is converted into a 1D binary vector, and the local structure surrounding an atom is encoded into a matrix that is the sum of the weighted vector representation of all neighboring atoms (equation 1).\\
To better explain OFM, let us take an example, $FeTi$ alloy with Body Center Cubic (BCC) crystal structure as shown in Figure 1. The center $Fe$ atom, denoted as $c$, is surrounded by fourteen neighbor atoms, denoted as $n$. The fourteen neighbor atoms include eight nearest $Ti$ atoms and six next-nearest $Fe$ atoms. Figure 1 shows the formation of a Voronoi polyhydron between the center $Fe$ atom and its fourteen neighbor atoms which has the shape of a truncated octahedron. The OFM consists of two parts: 1) a weight function $w_{cn}$ associated with center $Fe$ atom and any of these fourteen neighbor atoms (center-neighbor pair). $w_{cn}$ is calculated by multiplying $\theta_{cn}$, the solid angle subtended at the center atom by the face of the Voronoi cell corresponding to the neighbor atom, i.e., the solid angle between center-neighbor pairs in the Voronoi cell, with $\zeta(r_{cn})$ which is a function of the distance between them ($w_{cn}(r_{cn}) = \theta_{cn}\times \zeta(r_{cn})$). $\zeta(r_{cn})$ incorporates the information on the size of valence orbitals of the center-neighbor pairs as well as their interactions. 2) A 1D binary vector representation of each atom.  The electron configuration of valence orbitals for each atom is encoded into a 1D binary $32\times1$ vector (Figure 1). To construct the OFM for the center $Fe$ atom, we sum the matrix product for each center-neighbor pair and multiply it with the corresponding weight function. This results in a $32\times32$ matrix. Finally, to incorporate the information of the center $Fe$ atom, we concatenate its 1D binary vector to the $32\times32$ matrix, resulting in a $32\times33$ OFM for center $Fe$ atom.
Equation 1 describes the mathematical formulation of OFM. $X^c$ is the OFM for center $Fe$ atom, $M$ is the number of neighbor atoms surrounding the center $Fe$ atom,  $\overrightarrow O^c$ and $\overrightarrow O^n$ are the 1D binary vectors for the center $Fe$ and its fourteen neighbor atoms, respectively, and $\overrightarrow{O^c}^T$ is the transpose of $\overrightarrow{O^c}$. The incorporation of geometry as weight enables to encode the local chemical environments through orbital-orbital interactions. To investigate the efficiency of our network, we examined different $\zeta(r_{cn})$ functions including $\frac{1}{r_{cn}}$, $\frac{1}{r_{cn}^2}$, $\frac{1}{r_{cn}^4}$ and $\frac{1}{r_{cn}^6} - \frac{1}{r_{cn}^{12}}$. Subsequently, the $\zeta(r_{cn})$ function that resulted in best performance was selected. We note that in Voronoi polyhedra large solid angles correspond to shorter distance between two adjacent atoms and vice versa. In addition, such Voronoi polyhedra between the center and  neighbor atoms capture the geometry of the local environments of atoms. If only the distances between the center and the neighbor atoms are considered, the geometry will not be captured correctly.
\begin{figure}[H]
        \includegraphics[width=8.5cm,height=6.5cm]{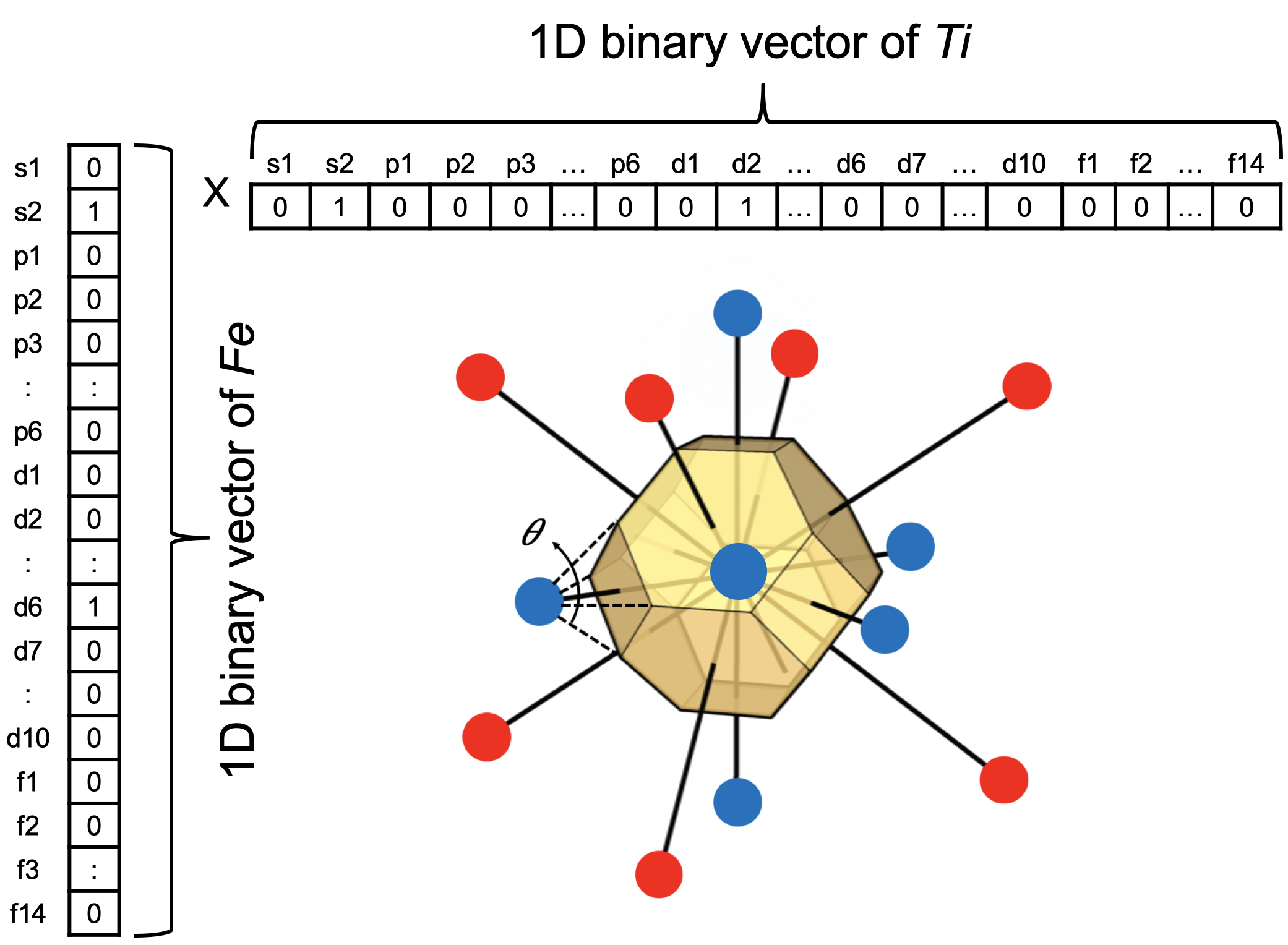}
        \label{fig:subim2}

\caption{OFM representation for $FeTi$ alloy.  Blue and red atoms are $Fe$ and $Ti$, respectively. The inset shows the Voronoi polyhedron for center $Fe$ atom forming a truncated octahedron.  The 1D binary vectors for $Fe$ and $Ti$ atoms are shown as well.}
\end{figure}
\vspace{-1cm}
\begin{equation}\label{eq:pareto mle2}
\begin{aligned}
X^c = \overrightarrow{O^c}^T + \sum_{n=1}^{M} \overrightarrow{O^c}^T \times \overrightarrow{O}^n \times {\theta_{cn}}\times{\zeta(r_{cn})}
\end{aligned}
\end{equation}
\noindent
\\
\noindent
Once we convert the orbital-orbital interactions between each atom and its neighbors into OFM representation, we use the graph convolutional neural network (GCNN) and couple it with OFM. We call this new model the OGCNN. The OGCNN network can be considered as a combination of four modules (Figure 2). The first module, input module, takes the basic atomic and the OFM features. In this module, the embeddings for all crystals in a batch are generated. The basic atomic features include properties like the group number, period number, and electronegativity. The list of 92 basic atomic features are provided in the Supplemental Material. To include the OFM features in the input module, the $32\times33$ OFM features corresponding to each atom are flattened into a $1056\times1$ vector. The atom features are constructed by combining the $92\times1$ basic atomic features and 1056 OFM features forming a $1148\times1$ vector for each atom in the crystal generating a unique representation for all crystals (equation 2).

\begin{equation}\label{eq:pareto mle2}
\begin{aligned}
v_i = [v_{oi}]+[v_{ai}]
\end{aligned}
\end{equation}
where $v_{oi}$ and $v_{ai}$ are the OFM and basic atomic features, respectively. The second module, the Encoder-Decoder module learns important features among atom features by employing a multilayer perceptron (MLP) with two fully connected layers. The MLP acts as an encoder-decoder network which comprises of 1148 at the encoder input, 768 neurons in the hidden layer and 1536 neurons at decoder output. Using such an architecture enables the network to select the most significant features among the pool of atomic and OFM features (Figure 2).
\begin{figure*}[t]
   \centering
\includegraphics[width=17 cm, height =14cm]{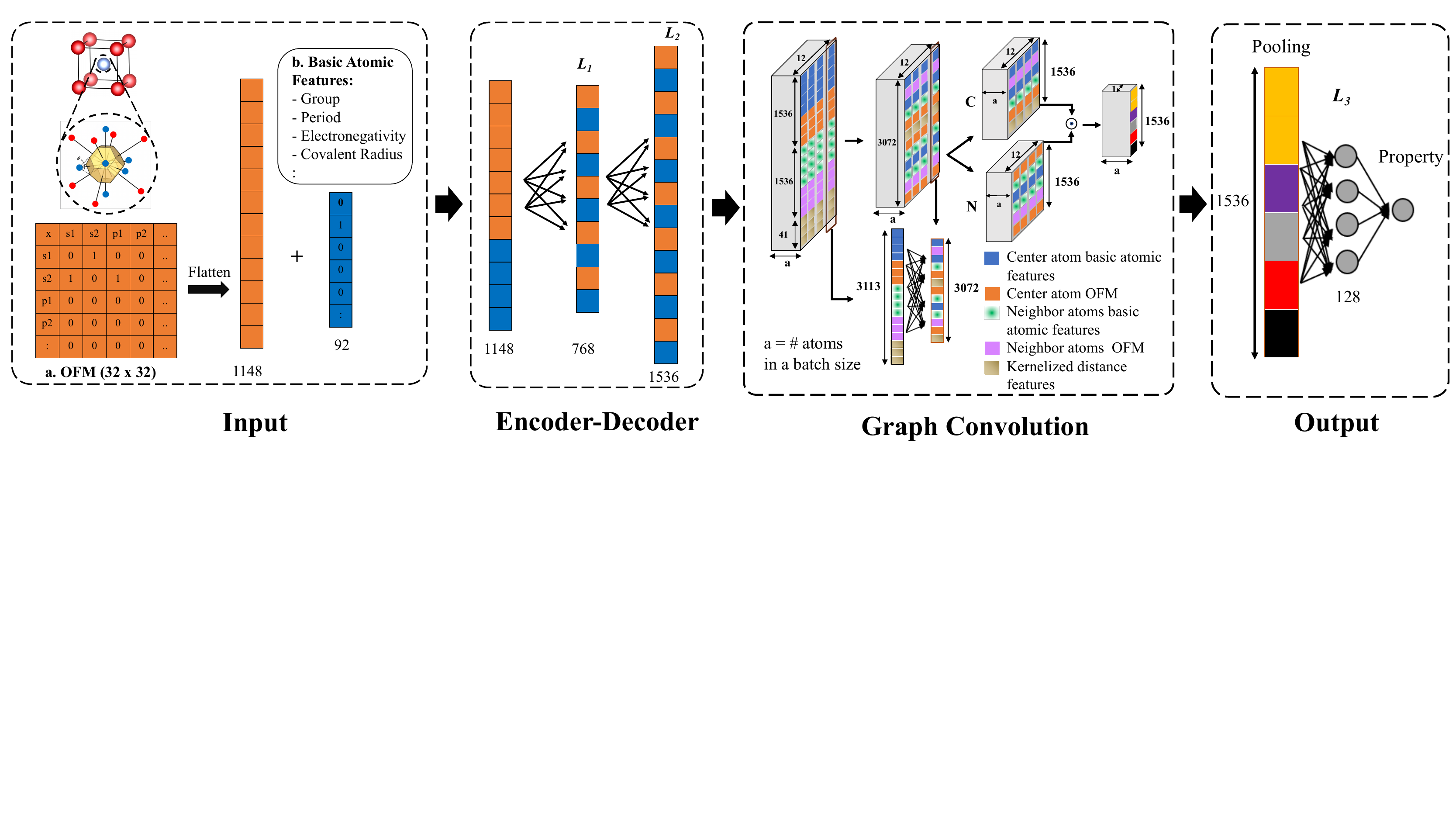}
\vspace{-55mm}
\caption{The structure of the OGCNN framework. It can be divided into four modules, the Input, Encoder-Decoder, Graph Convolution and output modules.}
\end{figure*}

In this paper, we consider the crystal structure as graph \cite{PhysRevLett.120.145301}. A crystal graph is an undirected graph in which the atoms are considered as nodes and the bonds as the edges, and each node has combination of basic atomic and OFM features (equation 2). This atom feature vector is then transformed as explained in Encoder-Decoder module to generate $V_i$ that includes the most important and relevant features. An important aspect of the crystal graph is that different atoms can be connected with more than one bond indicating multiple edges among the nodes of the graph. To incorporate the influence of neighbor atoms, the kernelized distance features between the $i$th and $j$th atoms are captured by the vector $u_{(i,j)_k}$ where k indicates the $k$th bond between them. The convolution operation is then performed as in Ref. \cite{PhysRevLett.120.145301},

\begin{equation}
    V_{i}^{(t+1)} = V_{i}^{(t)} + \sum \sigma(z^{(t)}_{(i,j)_k}W_f^{(t)} + b^{(t)}_f) \odot g(z^{(t)}_{(i,j)_k}W_s^{(t)} + b^{(t)}_s)
\end{equation}
where $z_{(i,j)} = V_i \bigoplus V_j \bigoplus u_{(i,j)_k} $, $\sigma$ is sigmoid activation function, and $g$ is the softplus function \cite{NIPS2000_1920}. $W$ and $b$ indicate the weights and bias in the network, respectively.\\ 
The convolution operation is performed three times in OGCNN network. Subsequently, a summation operation is performed over all the neighbors to aggregate the contribution of neighbor atoms which is then sent to the output module. Finally, in the  output module, a pooling operation is performed on the output from the graph convolution module to map the properties to a crystal level . The output of the pooling layer is subsequently used to predict the target property via a fully connected network with two layers.
More details about the architecture of the OGCNN model and the different hyperparameters optimized for training the network are available in the Supplemental Material.





\section{Training and Results}

To train the OGCNN model, we used five different DFT calculated datasets that include a diverse set of inorganic-crystals ranging from metals to complex minerals and oxides \cite{doi:10.1063/1.4812323,C2EE22341D}. The details about these datasets are provided in the Supplemental Material. To examine the generality of the model for predicting a wide variety of properties, we trained the OGCNN model for different properties including formation energy, band gap, and Fermi energy. For training the OGCNN model, we used mean square loss (MSE) as a loss function and stochastic gradient descent (SGD) as an optimizer. Additionally, for all cases, the entire datasets were split into $80$, $10$, and $10\%$ for training, testing and validation, respectively. Moreover, to avoid any bias during the training process, a five-fold cross-validation is used to split the datasets, and only their average values are reported \cite{Refaeilzadeh2009}.It must be noted that the OGCNN model was trained for 100 epochs and the weights of the model where lowest validation error was observed were used to predict the properties of crystals in the test set. The results for OGCNN and other models on the test set are summarized in Figure 3 and Table 1.
Figure 3(a) shows that for all properties and datasets the MAE values when using OGCNN are significantly lower compared to that of CGCNN. Figure 3(b) shows the percentage improvement for prediction accuracy using the OGCNN over CGCNN. The highest improvement in performance was achieved for the Lanthanides dataset: the OGCNN yields a MAE value of 0.061 eV/atom, whereas the CGCNN yields $0.133$ eV/atom. This corresponds to $54\%$ improvement in accuracy prediction for the OGCNN over CGCNN. On the other hand, the lowest performance over the CGCNN was achieved for the prediction of the Fermi energies of crystals from Materials Project (MP-Fermi energy) that is 0.38 eV corresponding to an improvement of $11\%$ over the CGCNN. Similarly, for other property and datasets including formation energies of Perovskites, formation energies of crystals from Materials Project (MP-formation energy), and band gaps of crystals from Materials Project (MP-band gap), a reduction of $50\%$, $45\%$ and $25\%$, respectively, in the MAE values were obtained using the OGCNN over that of the CGCNN.
To further benchmark the OGCNN, we compared the performance of the OGCNN in training Lanthanides dataset with some other material representations. For Lanthanides dataset, using OFM and Coulomb matrix representations and by using kernel ridge regression (KRR), the MAE values of 0.11 and 0.30 eV/atom have been reported \cite{PhysRevLett.108.058301,doi:10.1080/14686996.2017.1378060}.The CM representation has the lowest performance with MAE of 0.39 eV/atom followed by CGCNN, OFM, and OGCNN with MAE values of 0.13, 0.11, and 0.06 eV/atom, respectively (Figure 3(c)). The predicted formation energies of 741 test entries in Lanthanides dataset using the OGCNN against the DFT calculated values is also shown in Figure 3(d).
We also benchmarked our results against two previously developed state-of-the-art descriptors for encoding atomic structures including MBTR and SOAP \cite{huo2017unified,PhysRevB.87.184115}. Further details about the MBTR and SOAP hyperparameters optimization can be found in the the Supplemental Material.

\begin{figure}[H]
        \includegraphics[width=8cm,height=8.08cm]{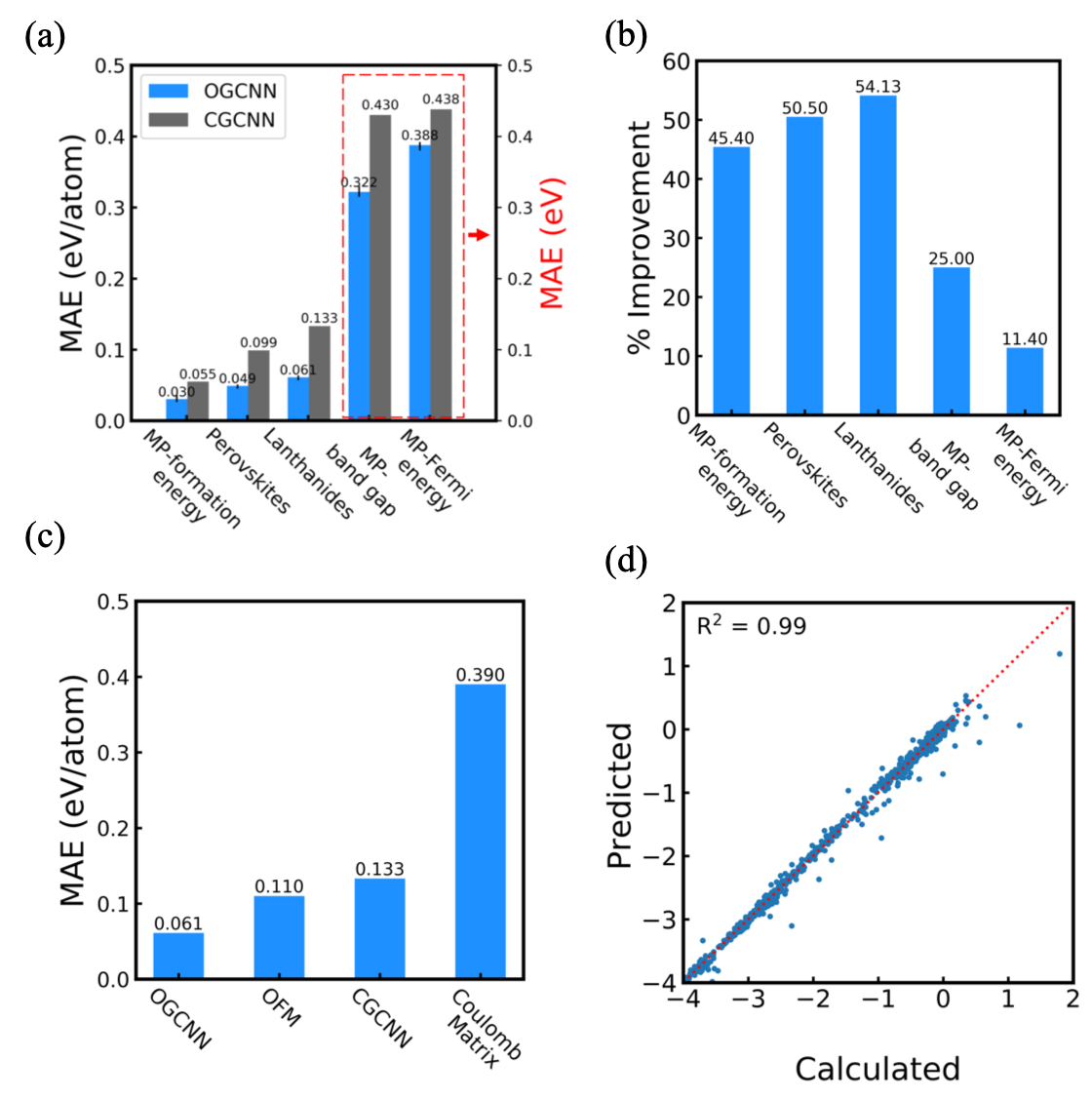}
        \label{fig:subim2}
\caption{Test set performance of OGCNN in different properties predictions across classes of material. (a) comparison between MAE values of OGCNN and CGCNN on test sets for five different datasets. (b) Percentage reduction of MAE values in (a) when OGCNN method is used for training over CGCNN method. Similar train-validation-test ratio was chosen for OGCNN and CGCNN networks. (c) MAE values for prediction of formation energies of Lanthanides dataset using different crystal representations. (d) Comparison of predicted formation energies of Lanthanides using OGCNN against DFT calculated values.}

\label{fig:image2}
\end{figure}
In Table 1, the performance of the OGCNN with CGCNN, SOAP and MBTR descriptors when used to predict different properties corresponding to five datasets were compared. We found that OGCNN exhibits highest performance among all. The only exception is the Fermi energy property, where both OGCNN and SOAP performed equally well.
We would like to emphasize that the MAE values for some properties in this study are within a narrow range from DFT calculated values (Table S4 in the Supplemental Material). For instance, the formation energy predictions using the OGCNN model for different datasets are within a range of 0.03-0.06 eV/atom from DFT calculated values. The MAE value for DFT calculations for formation energy with respect to experimental measurements is within the range of 0.081–0.136 eV/atom. Moreover, the desired chemical accuracy for formation energy is of order 0.04 eV/atom \cite{doi:10.1063/1.4812323, ONG2013314}. Similarly, for band gap property, a MAE value of the 0.6 eV for DFT calculations has been reported, and using OGCNN we obtained a MAE value of 0.32 eV \cite{Kirklin2015}.  Therefore, given the relatively low error in the predicted properties using the OGCNN, it can be reliably used to predict properties of new materials.

\renewcommand{\thetable}{\arabic{table}}
\begin{table}[H]

\caption {The mean absolute error values for test sets of five different datasets with OGCNN and CGCNN have been compared with the ones using SOAP and MBTR material representations.  }

\begin{tabular}{llllll}
 Dataset & \multicolumn{4}{c}{Material Representations}
\cr

        &  OGCNN  & CGCNN &  SOAP    &  MBTR                      
  \\    \hline
 Lanthanides \\ formation energy \\ unit - $ev/atom$   &  0.06    & 0.13    &
 0.09   & 0.28 
  \\
  \\
 Perovskites \\ formation energy \\ unit - $ev/atom$    &  0.05    & 0.09  &0.11 &0.09  
         \\  
         \\
 MP\\formation energy \\ unit - $ev/atom$   &  0.03  & 0.05 &0.05 &20 
  \\
  \\
 MP\\band gap \\ unit  - $ev$    &  0.32    & 0.43 &0.33 &0.69      
 \\  
 \\
 MP\\Fermi energy \\ unit  - $ev$  &0.38 &0.43 &0.38 &0.82 
 \\
 
\end{tabular}

\end{table}


\section{Conclusion}
In this study, we proposed  orbital crystal graph convolutional neural network, referred to as the OGCNN, which embeds atomic orbital-orbital interactions features  and basic atomic features. The OGCNN then was applied to different datasets to predict a wide range of material properties of versatile structures. The prediction accuracy of the OGCNN model is significantly higher than that of previously reported models. The inclusion of orbital-orbital interactions to encode the local chemical environments, and using encoder-decoder network were the fundamental reasons behind superior performance of the OGCNN. We expect this model to be applicable to a broader range of material discovery applications. The Github Repository for the OGCNN can be found at Ref. \cite{CodeRepo}.

\section*{Acknowledgements}

The template for the preprint has been taken from: \url{https://github.com/brenhinkeller/preprint-template.tex}

\normalsize
\bibliography{}



\end{document}